\documentclass[%
twocolumn,
showpacs,
bibnotes,
amsmath,amssymb,
aps,
prb,
]{revtex4-1}

\usepackage{epsfig}
\usepackage{graphicx}% Include figure files
\usepackage{dcolumn}% Align table columns on decimal point
\usepackage{bm}% bold math
\usepackage{color}

\usepackage{multirow}

\begin{document}

%\preprint{APS/123-QED}

\title{Role of $d$ orbitals in the Rashba-type spin splitting for noble-metal surfaces}% Force line breaks with \\

\author{Hyungjun Lee}
\affiliation{Department of Physics and IPAP, Yonsei University, Seoul 120-749, Korea.}%Lines break automatically or can be forced with \\
\author{Hyoung Joon Choi}
%\email[To whom correspondence should be addressed. ]{h.j.choi@yonsei.ac.kr}
\email{h.j.choi@yonsei.ac.kr}
\affiliation{Department of Physics and IPAP, Yonsei University, Seoul 120-749, Korea.}%Lines break automatically or can be forced with \\
%\affiliation[Also at ]%{%


%  \homepage{http://www.Second.institution.edu/~Charlie.Author}
% \affiliation{
%  Second institution and/or address\\
%  This line break forced% with \\
% }%
% \affiliation{
%  Third institution, the second for Charlie Author
% \affiliation{%
%  Authors' institution and/or address\\
%  This line break forced with \textbackslash\textbackslash
% }%

% \collaboration{CLEO Collaboration}%\noaffiliation

\date{\today}% It is always \today, today,

\begin{abstract}
We investigate the Rashba-type spin splitting in the Shockley surface states 
on Au(111) and Ag(111) surfaces, based on first-principles calculations. 
By turning on and off spin-orbit interaction (SOI) partly, we show that 
although the surface states are mainly of $p$-orbital character with only 
small $d$-orbital one, $d$-channel SOI determines the splitting 
and the spin direction while $p$-channel SOI has minor and 
negative effects. The small $d$-orbital character of the surface states, 
present even without SOI, varies linearly with the crystal momentum $k$, 
resulting in the linear $k$ dependence of the splitting, the Hallmark of 
the Rashba type. As a way to perturb the $d$-orbital character of the surface 
states, we discuss effects of electron and hole doping to the Au(111) surface.

\end{abstract}


% 73.20.At : Surface states, band structure, electron density of states
% 71.70.Ej : Spin-orbit coupling, Zeeman and Stark splitting, Jahn-Teller effect
% 79.60.Bm : Clean metal, semiconductor, and insulator surfaces
% 71.18.+y : Fermi surface: calculations and measurements; effective mass, g factor


\maketitle

\section{Introduction}

Approximated to the second order in power of the inverse of the speed
of light, $c$, the relativistic Dirac equation is reduced to the usual
nonrelativistic Schr\"{o}dinger equation for the two-component spinor wave
function with the three relativistic correction terms: the mass-velocity 
term, the Darwin term, and the spin-orbit interaction (SOI) term.\cite{1} 
Among them, SOI induces many intriguing magnetic and electronic phenomena 
in condensed matter physics, such as the intrinsic spin Hall effect\cite{2,3} 
and the Rashba-Bychkov effect (simply, the Rashba effect) in two-dimensional 
(2D) semiconductor heterostructures,\cite{4} and recently it is known as 
one of the main mechanisms of the generation of the robust metallic surface 
(or edge) states in topological insulators.\cite{5} These phenomena have been 
expected to open a new pathway to the spin-dependent electronics and the
quantum computing, and thus SOI in condensed matters has been of great 
interest both theoretically and experimentally.\cite{5,6}

So far, exploration of SOI phenomena has been limited to materials with 
elements of high atomic numbers such as noble metals of Pt and Au, and $p$-orbital 
(semi-)metals like Sb, Tl, Pb, and Bi. Among them, Au has the atomic SOI 
strength, $(1/2)\xi_{5d}$, of 0.35~eV for the $5d$ level,\cite{7,8}
attracting special interest due to the SOI-induced Rashba-type spin-splitting 
of the surface states in Au(111) surface first observed by angle-resolved 
photoemission spectroscopy.\cite{9} Such spin splitting 
was also observed in other clean (semi-)metallic surfaces \cite{10,11} 
and surface alloys;\cite{12,13} however, the splitting was unresolvably 
small in another noble-metal Ag surface\cite{14,15} although 
Ag has the atomic SOI strength, $(1/2)\xi_{4d}$, of 0.13~eV for 
the $4d$ level.\cite{7}

Previous theoretical studies pointed out that the dissimilarity in Au
and Ag is attributed to their differences in the intrinsic atomic SOI
strength\cite{16} and the degree of asymmetry of the wave function near the
nucleus of surface atoms or, stated differently, the hybridization of
different-parity orbitals in the surface-state wave functions.\cite{17,18} 
Compared with Ag, Au has a distinguishing feature that
the energy-level separation between $5d$ and $6s$ orbitals is rather
small,\cite{18,19,20} which is related to the relativistic mass-velocity and
Darwin terms.\cite{19} This proximity of $5d$- and $6s$-orbital energy levels in
Au increases $d$-orbital character in the surface states,\cite{18} which is found 
important for the splitting.\cite{17,18} However, the role of $d$ 
orbitals in the splitting is still unknown except that they make
the surface states asymmetric.

In our present work, we investigate the origin of the Rashba-type splitting 
in the Shockley surface states on Au(111) and Ag(111) surfaces, focusing 
on the role of $d$ orbitals. By turning on and off parts of the SOI 
Hamiltonian, we find that the splitting originates from interaction 
between $d$ orbitals via $d$-channel SOI although the surface states are 
derived mainly from $p$ orbitals with only small $d$-orbital character, 
and the spin direction is also determined by the $d$-channel SOI. The 
linear dependence of the splitting on the wave vector, $k$, which is the 
Hallmark of the Rashba type, is found to be derived from $k$-linear dependence of 
$d_{xz}\,(d_{yz})$-orbital character of the surface states which exists 
even without SOI. These results show that the small $d$-orbital character 
of the noble-metal surface states determines the Rashba-type splitting 
while the large $p$-orbital one has weak and negative effects.
As a way to modify the splitting by varying $d$-orbital character 
of the surface states, we consider electron and
hole doping to the Au(111) surface.

\section{Calculational method}

Our present calculations are based on \textit{ab initio} density-functional
methods,\cite{21,22} which employ a Troullier-Martins norm-conserving
pseudopotential\cite{23} and the Perdew-Burke-Ernzerhof-type
generalized-gradient approximation for the exchange-correlation
potential.\cite{24} Wave functions are expanded with the localized
pseudoatomic orbital bases.\cite{22} The SOI term is incorporated within a
fully relativistic $j$-dependent pseudopotential,\cite{25} and treated in the
$l$-dependent fully-separable nonlocal form using additional Kleinman-Bylander 
projectors.\cite{26,27} In treating the SOI Hamiltonian, we consider 
non-self-consistently only the effect of atomic core potentials
with the $\vec{L}\cdot\vec{S}$ form, where $\vec{L}$ and $\vec{S}$
are the orbital and spin angular momentum operators, respectively. 
Although we can also include contribution of self-consistent part of
potentials to SOI, our test calculations confirm that this contribution 
makes very little difference in the electronic band structures\cite{17} 
which is smaller than 10$^{-4}$~eV. Thus, we do not include the contribution of 
the self-consistent part of the potentials to SOI in our following calculations.

For surface calculations, we model Au(111) and Ag(111) surfaces with 
supercell including a slab of 30 atomic layers and a vacuum region of 
about 20 {\AA} thickness, and we take 36$\times$36$\times$1 
Monkhorst-Pack special-$k$-point meshes in the full Brillouin zone 
(BZ)\cite{28} to integrate the charge density. While Ag(111) exhibits no
surface reconstruction, the Au(111) surface reportedly has a surface
reconstruction of a 22$\times\sqrt{3}$ unit cell, the ``herringbone
reconstruction''.\cite{29} Since this reconstruction has negligible 
effect on the electronic structures of the surface states,\cite{30} 
we safely take the clean-cut bulk atomic structures for Au(111) and 
Ag(111) surfaces.\cite{31}

\section{Electronic band structures for A\lowercase{u}(111) and
  A\lowercase{g}(111) surfaces}

\begin{figure}
  \centering
\epsfig{file=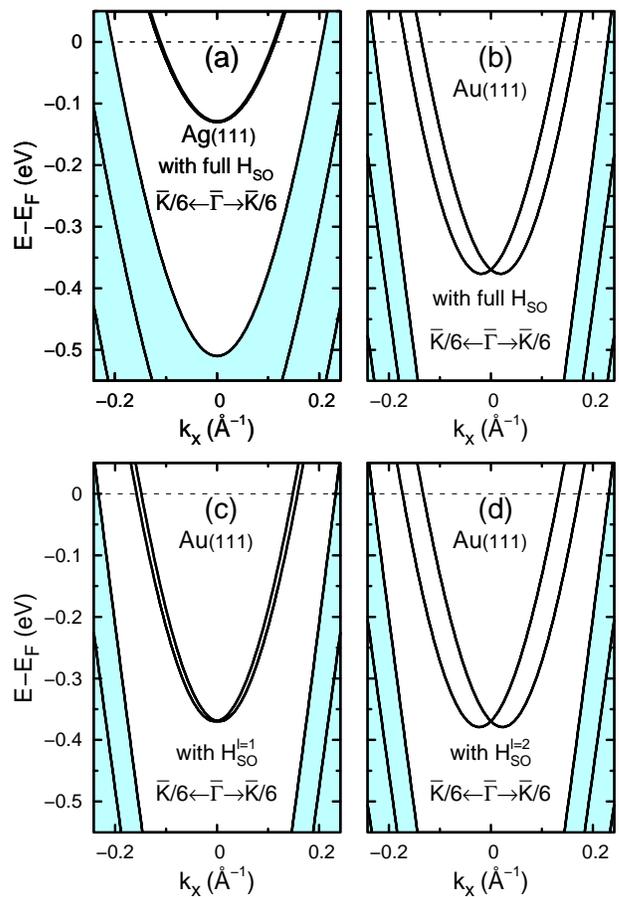,width=0.45\textwidth,angle=0,clip=}
  \caption{%
(Color online) Electronic band structures of (a) Ag(111) and (b) Au(111) 
surfaces around the 2D BZ center, $\bar{\Gamma}$, calculated with the full SOI,
$H_\mathrm{SO}(k)$, and those
of Au(111) surface calculated (c) with only $p$-channel SOI, 
$H_\mathrm{SO}^{l=1}(k)$, and (d) with only $d$-channel SOI, 
$H_\mathrm{SO}^{l=2}(k)$. The Fermi energy is set to zero. 
The bulk-projected electronic states are shaded in cyan.}
\end{figure}

Figures~1(a) and (b) show our calculated band structures of Ag(111) and 
Au(111) surfaces, respectively. Compared with corresponding experimental 
values,\cite{14} the maximum binding energies of our surface bands 
are 60~meV larger for Ag(111) [Fig.~1(a)] and 100~meV smaller for Au(111) 
[Fig.~1(b)], consistent with previous all-electron calculations.\cite{15}
The accuracy of the binding energy can be improved by careful description of 
decay of the wave functions into the vacuum,\cite{32} but we already have 
good accuracy for the Rashba-type splitting. Our surface bands show 
$k_0 = 0.013$~{\AA}$^{-1}$ for the Rashba splitting in $k$ space, 
defined by the offset by which the extremum of the surface-state dispersion 
shifts from the crossing point $\bar{\Gamma}$ and $E_\mathrm{R} = 2.8$~meV 
for the Rashba energy, defined by the energy difference between the crossing 
point and the maximum binding energy. When we fit our surface bands to an 
expression, $E_\pm=\frac{\hbar^2k^2}{2m^*}\pm\alpha k+E_0$, we obtain the
electron effective mass $m^*=0.26$~$m_\mathrm{e}$, where $m_\mathrm{e}$ is the 
electron bare mass, and the Rashba parameter $\alpha$ = 0.460~eV$\cdot${\AA}. 
These results are in good agreement with the reported 
experiments\cite{9,14,15} and theories.\cite{15,18,31}
In addition, if we shift the surface bands to match the maximum binding 
energy with the experimental value, $487\,\mathrm{meV}$ for Au(111), the 
calculated Fermi wave vectors are 0.165 and 0.194~{\AA}$^{-1}$ for the inner 
and outer Fermi surfaces (FSs), respectively, which are also in good 
agreement with reported experimental and theoretical results.\cite{14,15}
These results show that our Kohn-Sham (KS) Hamiltonian, 
$H_\mathrm{KS}(k)=H_0(k)+H_\mathrm{SO}(k)$, which is the sum of the scalar
relativistic part, $H_0(k)$, and the SOI term, $H_\mathrm{SO}(k)$, 
describes the Rashba-type splitting in Au(111) surface accurately.

\section{Orbital contributions to the Rashba splitting in the Surface States on A\lowercase{u}(111) surface}

In semiconductor devices, the relativistic response of electron spin to 
the electric field, $E\hat{n}$, yields the $k$-linear Rashba 
Hamiltonian,\cite{4}
$H_\mathrm{R}=\alpha_\mathrm{R}(\hat{n}\times\vec{k})\cdot\vec{\sigma}$,
with the Pauli matrices $\vec{\sigma}$ and
$\alpha_\mathrm{R}=e\hbar^2E/(4m_\mathrm{e}^2c^2)$. These terms are
derived from the microscopic SOI Hamiltonian
\begin{equation}
  H_\mathrm{SO}=\frac{\hbar}{4m_\mathrm{e}^2c^2}
                (\nabla V\times\vec{p})\cdot\vec{\sigma}
\end{equation}
and the nearly-free electron nature for 2D electron gases confined to 
surfaces or interfaces. While this model can explain the $k$-linear Rashba 
splitting, it cannot explain the correct size of splitting\cite{9,16} due to
ignoring the effect of ion cores.\cite{9}

To find the origin of the Rashba-type splitting, we consider 
orbital-angular-momentum ($l$) decomposition of the SOI term, i.e., 
$H_\mathrm{SO}(k)$ = $H_\mathrm{SO}^{l=1}(k)$ + $H_\mathrm{SO}^{l=2}(k)$,
and calculate the band structures with either $H_\mathrm{SO}^{l=1}(k)$ or 
$H_\mathrm{SO}^{l=2}(k)$. This approach is possible because 
$H_\mathrm{SO}(k)$ in our calculation is
expressed as a sum of $l$-dependent Kleinman-Bylander-type
projectors. Figures~1(c) and (d) show that when only the $p$-channel SOI, i.e.,
$H_\mathrm{SO}^{l=1}(k)$, is switched on, the band splitting of the surface 
states is reduced greatly, but when only the $d$-channel SOI, i.e.,
$H_\mathrm{SO}^{l=2}(k)$, is switched on, the splitting is even slightly larger 
than that for the full $H_\mathrm{SO}$ case. While $m^*$ in both cases are
very close to the full $H_\mathrm{SO}$ case, the Rashba parameter 
is $\alpha_p$ = 0.118 eV$\cdot${\AA} in the former case and $\alpha_d$ 
= 0.564 eV$\cdot${\AA} in the latter case, giving the
  following relation,\cite{comm}
\begin{equation}
  \alpha_\mathrm{full} \approx -\alpha_p+\alpha_d~.
\end{equation}
This shows that the Rashba-type splitting in Au(111) surface originates 
from the $d$-channel SOI while the $p$-channel SOI slightly reduces the 
splitting.

In the above, we examined $l$ decomposition of
the SOI Hamiltonian. Now we consider orbital decomposition of the
surface states on Au(111) surface, i.e., 
$\psi_{nk}=\psi_{nk,s}+\psi_{nk,p}+\psi_{nk,d}$, where $\psi_{nk}$ 
is the wave function of the $n$th surface band for the wave vector $k$ and 
$\psi_{nk,s}$, $\psi_{nk,p}$, and $\psi_{nk,d}$ are $s$-, $p$-, and 
$d$-orbital parts of the state, respectively.
With this decomposition, we calculate the expectation values 
$\langle\psi_{nk,i}|H_\mathrm{SO}(k)|\psi_{nk,j}\rangle$ 
of the full $H_\mathrm{SO}$ for $i,\,j=s,\,p,\,d$ to analyze 
orbital contributions to the Rashba-type splitting. We find that
the $d$-$d$ SOI energy, i.e.,
$\langle\psi_{nk,d}|H_\mathrm{SO}(k)|\psi_{nk,d}\rangle$, has opposite signs 
for the two bands, contributing most substantially to the band splitting.
Then, the $p$-$d$ SOI energy, i.e.,
$\langle\psi_{nk,p}|H_\mathrm{SO}(k)|\psi_{nk,d}\rangle$, contributes
to the splitting by about 10~\% of the $d$-$d$ one. In contrast, the $p$-$p$ and 
$s$-$p$ SOI energies contribute weakly and negatively to the band 
splitting, i.e., they reduce the splitting slightly, and the $s$-$s$
and $s$-$d$ ones are almost zero. These results again show that the 
$d$-orbital character of 
the surface states determines the Rashba-type splitting in Au(111) surface 
although the surface states are mainly of $p$-orbital one.

\begin{figure}
\centering
\epsfig{file=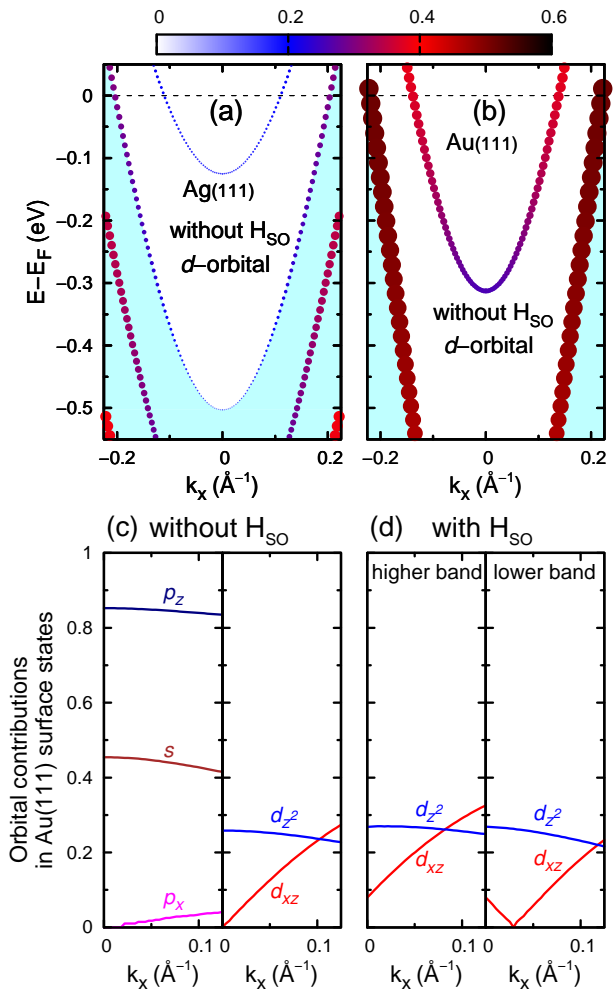,width=0.45\textwidth,angle=0,clip=}
\caption{%
 (Color online) Orbital characters of Au(111) and Ag(111) surfaces.
 (a, b) Band structures without SOI for (a) Ag(111) and (b) Au(111) 
 surfaces. Both the color and the size of filled dots represent the norm
 of all $d$-orbital characters ($|\psi_{n\vec{k},\mathrm{all}\, d}|$) of 
 each wave function. The norm is defined using the pseudoatomic 
 orbitals.\cite{norm} The bulk-projected electronic states are shaded in cyan.
 (c, d) Orbital characters (c) without SOI and (d) with SOI for Au(111) 
 surface states along the positive $k_x$ axis. In (c), norms of $s$-, $p_x$-, 
 and $p_z$-orbital characters ($|\psi_{n\vec{k},s}|$, $|\psi_{n\vec{k},p_x}|$, 
 and $|\psi_{n\vec{k},p_z}|$) are in the left panel, and norms of $d_{xz}$- 
 and $d_{z^2}$-orbital characters ($|\psi_{n\vec{k},d_{xz}}|$ and 
 $|\psi_{n\vec{k},d_{z^2}}|$) are in the right panel. In (d), norms of 
 $d_{xz}$- and $d_{z^2}$-orbital characters ($|\psi_{n\vec{k},d_{xz}}|$ and 
 $|\psi_{n\vec{k},d_{z^2}}|$) of the higher- and lower-energy surface states 
 are in the left and right panels, respectively. The norm, 
 $|\psi_{n\vec{k},\mu}|$, of an orbital character, $\mu$, is defined
 using the pseudoatomic orbitals.\cite{norm}
}
\end{figure}

To verify this picture for the difference in the Rashba-type splitting in 
Au(111) and Ag(111) surfaces, we calculate the band structures without SOI 
and estimate $d$-orbital characters in their surface states,\cite{norm} 
similarly to the previous works.\cite{17,18} As shown in Figs.~2(a) and (b), 
the Ag(111) surface states have negligible $d$-orbital part, but the Au(111) 
surface states have relatively significant $d$-orbital one.\cite{17,18} This 
difference, due to the smaller energy separation between Au $5d$ and $6s$ 
levels than that between Ag $4d$ and $5s$ levels,\cite{18} confirms the 
importance of $d$ orbitals in the Rashba-type splitting.\cite{17,18}

As a double check, we consider shift of the bulk $d$-band energies, 
that is, we modify the KS Hamiltonian matrix by adding $\delta E_d$ to 
its diagonal elements for $d$ orbitals after self-consistency, and then 
calculate the band structure. As shown in Table I, with no shift 
($\delta E_d=0$), the Au(111) surface bands are split by 
$\Delta_{k_\mathrm{F}}$  = 135~meV at the inner Fermi wave vector, and 
the Rashba splitting in $k$ space is $k_0 = 0.013$~{\AA}$^{-1}$, as 
mentioned above, and the Ag(111) surface bands have almost zero energy 
splitting (3~meV). For Au(111), when $\delta E_d=-$0.5~eV, the energy 
splitting $\Delta_{k_\mathrm{F}}$ decreases to 92~meV and $k_0$ decreases 
to 0.009~{\AA}$^{-1}$ and when $\delta E_d=$1.0~eV, 
$\Delta_{k_\mathrm{F}}$ increases to 191~meV and $k_0$ increases to 
0.019~{\AA}$^{-1}$. For Ag(111), when $\delta E_d=5.0$~eV, 
$\Delta_{k_\mathrm{F}}$ increases to 48~meV and $k_0$ increases to 
0.006~{\AA}$^{-1}$. These results show that the Rashba-type splitting in 
noble metals indeed depends on $d$-band energies. We also note that 
$\delta E_d=5.0$~eV makes the shifted $d$-band energy in Ag close to the 
unshifted $d$-band energy in Au, but the splitting in Ag(111) is still 
smaller than that in Au(111) due to the difference in the intrinsic 
atomic SOI strength.

\begin{table}[b]
  \setlength{\extrarowheight}{2pt}
  \setlength{\tabcolsep}{10pt}
  \newcolumntype{.}{D{.}{.}{-1}}
    \caption{Dependence of the Rashba splitting on the bulk $d$-band
      energy. Values of $\Delta_{k_\mathrm{F}}$ and $k_0$ are calculated with
      the shift of the bulk $d$-band energy by $\delta E_d$. For a
      better comparison, $\Delta_{k_\mathrm{F}}$ refers
      to the splitting at the same $k$ which is the inner Fermi wave
      vector in the case of $\delta E_d = 0$.}
  \begin{tabular}{c...}
    \hline
    \hline
    & \multicolumn{1}{c}{$\delta E_d$\,(eV)} & \multicolumn{1}{c}{$\Delta_{k_\mathrm{F}}$\,(meV)}  & \multicolumn{1}{c}{$k_0$\,({\AA}$^{-1}$)}\\
    \hline
    \multirow{4}*{Au(111)} & -0.5 & 92 & 0.009\\
    & 0.0 & 135 & 0.013\\
    & 0.5 & 172 & 0.016\\
    & 1.0 & 191 & 0.019\\
    \hline
    \multirow{4}*{Ag(111)} & 0.0 & 3 & 0.000\\
    & 2.0 & 18 & 0.003\\
    & 4.0 & 39 & 0.006\\            
    & 5.0 & 48 & 0.006\\
    \hline
    \hline
  \end{tabular}
\end{table}

Focusing back to Au(111), Fig.~2(b) shows that the $d$-orbital part of the 
surface bands grows monotonously as $k$ increases from $\bar{\Gamma}$. This 
suggests that even the linear $k$ dependence of the Rashba-type splitting 
may be determined by $k$-dependent size of the $d$-orbital character that 
exists without SOI. Thus, with more analysis, we find that only $s$, $p_z$, 
$d_{xz}$, and $d_{z^2}$ orbitals are significant for surface states along 
the $k_x$ axis, when the $z$ axis is the positive direction normal to
the top surface. As shown in 
Fig.~2(c), $d_{xz}$-orbital character increases from zero, as $k$ increases 
along the $k_x$ axis, while $s$-, $p_z$-, and $d_{z^2}$-orbital characters 
decrease slowly. Considering the symmetry, surface states along different 
$k$ direction, i.e., along the $k_y$ axis, have $d_{yz}$ orbital 
significantly instead of $d_{xz}$. When we use $\psi_{k,d_{xz}}^{(0)}$ and 
$\psi_{k,d_{z^2}}^{(0)}$ for $d_{xz}$- and $d_{z^2}$-orbital parts of a 
surface state $\psi_k^{(0)}$ (not including spin) without SOI, respectively, 
we obtain that the SOI-induced energy splitting $\Delta_k$ of the two surface 
bands is given by
\begin{eqnarray}
  \label{eq:1}
  \Delta_k &\approx& 2\left|\langle\psi_k^{(0)}|
            H_\mathrm{SO}^{\uparrow\downarrow}(k)|
            \psi_k^{(0)}\rangle\right|\nonumber\\
  &\approx&  2\left|\langle\psi_{k,d_{xz}}^{(0)}|
            H_\mathrm{SO}^{\uparrow\downarrow}(k)|
            \psi_{k,d_{z^2}}^{(0)}\rangle\!
            +\!\langle\psi_{k,d_{z^2}}^{(0)}|
            H_\mathrm{SO}^{\uparrow\downarrow}(k)|
           \psi_{k,d_{xz}}^{(0)}\rangle\right|\nonumber\\
  &\propto& \left|\psi_{k,d_{z^2}}^{(0)}\right|\left|
            \psi_{k,d_{xz}}^{(0)}\right| \propto \left|
            \psi_{k,d_{z^2}}^{(0)}\right||k|\,,
\end{eqnarray}
where $H_\mathrm{SO}^{\uparrow\downarrow}(k)$ is SOI between up- and 
down-spin states. Thus, the linearly $k$-dependent $\psi_{k,d_{xz}}^{(0)}$ 
makes the Rashba-type splitting $\Delta_k$ proportional to $|k|$. As SOI 
splits the two surface bands, the $d_{xz}$-orbital 
character increases in the higher-energy surface state while it decreases 
in the lower-energy one [Fig.~2(d)]. SOI redistributes $d$-orbital
characters in the surface states but it does not mix the surface states 
with bulk states significantly. We note that even without SOI, the
$d_{xz}$ and $d_{z^2}$ orbitals in the surface states along the $k_x$ axis form 
a nonzero $y$ component of the orbital angular momentum, which may be 
related to a recent study of the Rashba splitting.\cite{37}

\section{Spin and orbital angular momentum of A\lowercase{u}(111)
  surface states}

\begin{figure}[b]
  \centering
\epsfig{file=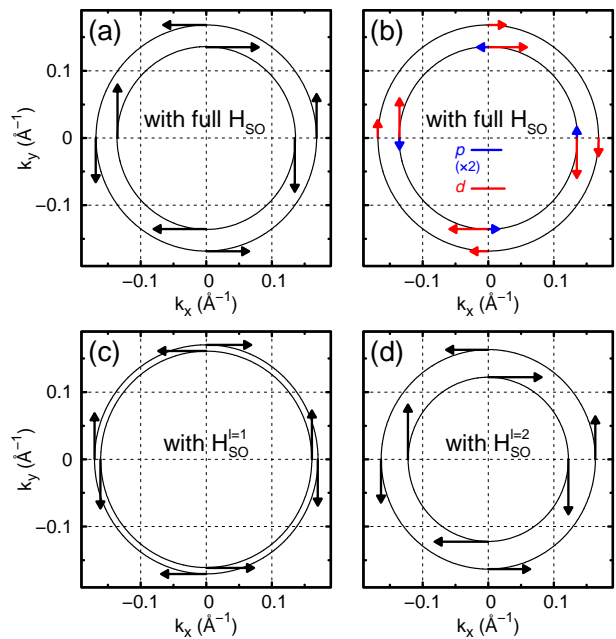,width=0.45\textwidth,angle=0,clip=}
  \caption{%
(Color online) SAM and OAM of Au(111) surface states at the Fermi energy. 
(a) SAM and (b) OAM with the full SOI.
(c) SAM  with only $p$-channel SOI, $H_\mathrm{SO}^{l=1}$. (d) SAM with 
only  $d$-channel SOI, $H_\mathrm{SO}^{l=2}$. Black arrows represent SAM,
and blue and red arrows do OAMs from $p$ and $d$ orbitals,
respectively. For clarity, OAM from $p$ orbital is doubled.}
\end{figure}

Another important feature of the Rashba-type splitting is the helical spin 
structure. To find the role of $d$ orbitals in the spin structure,
we calculate expectation values of the spin angular momentum (SAM) operator. 
In addition, motivated by a recent work on the topological
insulator,\cite{33} we also calculate the orbital angular momentum (OAM) by projecting the 
surface-state wave functions onto a set of localized functions defined in a 
sphere centered at the top-surface Au atom and then evaluating expectation 
values of OAM operators around the atom using the projected wave 
functions.\cite{33} A half of the nearest neighbor distance, 1.44 {\AA}, 
is used for the sphere radius, and real spherical harmonics multiplied 
by a constant radial part are used for the localized functions.

\begin{figure*}[t]
  \centering
\epsfig{file=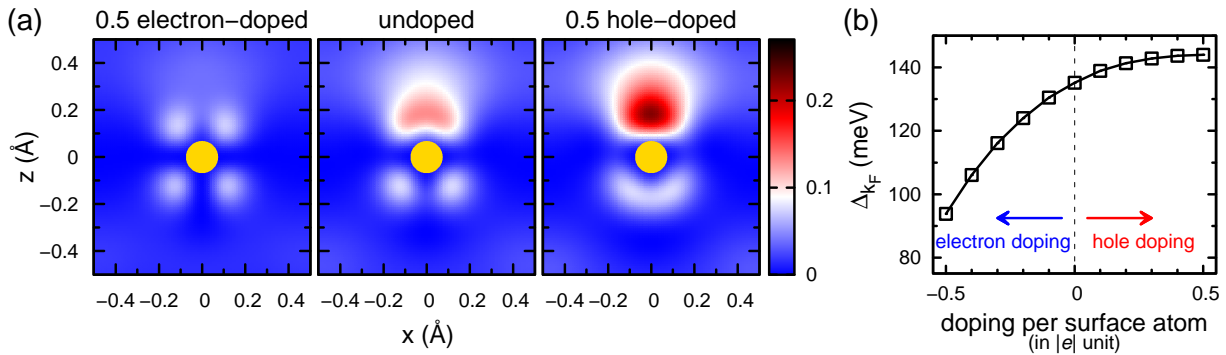,width=0.9\textwidth,angle=0,clip=}
  \caption{%
(Color online) Effects of doping on the Rashba-type splitting on Au(111) 
surface. 
(a) Squared amplitudes of surface-state wave functions at the same 
$k_\mathrm{F}$, the Fermi wave vector of the inner FS in the undoped 
case, in 0.5 electron-doped, undoped, 
and 0.5 hole-doped cases, shown from left to right panels. The color
scale is in units of per \AA$^3$.
The surface Au atom is marked with a dot at the center of each panel.
(b) Band splitting $\Delta_{k_\mathrm{F}}$ vs doping.  All splittings are taken at 
the same $k_\mathrm{F}$ as in (a).
}
\end{figure*}

As shown in Figs.~3(a) and (b), obtained SAM and OAM of the surface states 
are in-plane and helical around circular FSs. In our results, SAM is mainly 
from $p_z$ orbital while OAM is mainly from $d$ orbitals. When viewed 
from the vacuum side toward the top surface, SAM is clockwise at the 
inner FS and counter-clockwise at the outer FS [Fig.~3(a)], which is 
consistent with the reported experiments\cite{34} and
calculations.\cite{38} When viewed from the same side, OAM from $d$ orbitals 
is clockwise at both FSs while OAM from 
$p_z$ orbital is counter-clockwise at the inner FS and clockwise at the 
outer FS [Fig.~3(b)]. Total OAM is clockwise at both FSs and has smaller 
magnitude than SAM. These features for OAM are also in good agreement with 
the recent experimental and theoretical reports,\cite{38} but are different 
from the $p$-orbital-based Bi$_2$Se$_3$ topological insulator where OAM 
has opposite directions for the two surface bands and has larger amplitude 
than SAM.\cite{33}

We also study the roles of $p$- and $d$-channel SOIs in determining the spin
direction. As shown in Figs.~3(c) and (d), our result shows that when 
only the $p$-channel SOI, i.e., $H_\mathrm{SO}^{l=1}(k)$, is switched on, 
the spin direction is opposite to the full $H_\mathrm{SO}$ case given in 
Fig.~3(a), but when only the $d$-channel SOI, i.e., $H_\mathrm{SO}^{l=2}(k)$, 
is switched on, the spin direction is the same with the full $H_\mathrm{SO}$ 
case. This result shows again that the $d$-channel SOI is indeed decisive 
in the Rashba-type splitting while the $p$-channel SOI has minor and opposite 
effects.

\section{Electronic perturbations to A\lowercase{u}(111) surface}

As a way to change the Rashba-type splitting in the noble-metal surface,
we consider electron and hole doping, similarly to Xe and Na adsorption 
experiments.\cite{35} We performed self-consistent calculations with 
different numbers of electrons. In the undoped case, the surface-state 
wave function is asymmetric at the surface Au atom, with greater amplitude 
in the vacuum side than in the bulk side [Fig.~4(a)]. Compared with this, 
electron (hole) doping makes the wave function less (more) asymmetric 
[Fig.~4(a)], resulting from decrease (increase) of 
the $d_{z^2}$-orbital character in the wave function due to change 
in the degree of the inversion symmetry breaking, affected by change of
the potential near the surface. As shown in Eq.~\eqref{eq:1}, the Rashba
splitting $\Delta_k$ in noble-metal surfaces depends on the size of 
$d_{z^2}$ orbital in the surface-state wave functions. Thus,  
change in the $d$-orbital characters changes the energy splitting.
For Au (111) surface, we obtain 40 meV decrease for 0.5 electron doping per 
surface Au atom, and 10 meV increase for 0.5 hole doping [Fig.~4(b)]. 
These changes are $\sim$100 times the Rashba effect in semiconductor 
devices at electric fields corresponding to the doping concentrations. 

\section{Conclusion}

In conclusion, we investigated the Rashba-type splitting in the Shockley 
surface states on Au(111) and Ag(111) surfaces, based on first-principles 
calculations including SOI. We showed that although the surface states 
have predominantly $p$-orbital character and only small $d$-orbital one, 
interaction between $d$-orbital parts via the $d$-channel SOI produces
the Rashba-type splitting. The small $d_{xz}\,(d_{yz})$-orbital part of 
the surface states, present even without SOI, depends linearly on $k$ so 
that the atomic SOI results in the linearly $k$-dependent Rashba-type 
splitting. The $d$-channel SOI determines the helical spin structure of 
the surface states although the spins are mainly from $p$ orbitals. The 
$p$-channel SOI has minor and opposite effects in Au(111) surface. Thus, 
the role of $d$ orbitals in Au(111) surface has some similarity to graphene 
where small $3d$ orbitals contribute dominantly to the SOI-induced energy 
gap,\cite{36} but shows difference from $p$-orbital-based Bi$_2$Se$_3$-type
topological insulators.\cite{33} As a way of varying the splitting size 
in Au(111) surface, we discussed effects of electron and hole doping.

\begin{acknowledgments}
This work was supported by the NRF of Korea (Grant No. 2011-0018306).
Computational resources are provided by KISTI Supercomputing Center 
(Project No. KSC-2011-C3-05).
\end{acknowledgments}

\nocite{*}


\begin{thebibliography}{30}
\bibitem{8} The atomic SOI strength, $\xi$, of an atomic orbital
  is
  \begin{equation*}
    \xi=\int\frac{\hbar^2}{2m_\mathrm{e}^2c^2}\frac{1}{r}\frac{dV(r)}{dr}[R(r)]^2r^2\,dr\,,
  \end{equation*}
  where $m_\mathrm{e}$ is an electron mass, $V(r)$ is a potential, and $R(r)$ is the radial part of the atomic orbital.
\bibitem{comm} 
Examining the coefficients of pseudoatomic orbitals in the surface-state 
wave functions, we find that the negative contribution from the $p$-channel 
SOI with respect to the $d$-channel SOI in Eq.~(2) results from the opposite 
relation of the relative phases in the on-site hybridizations between 
$p_x$ ($p_y$) and $p_z$ orbitals and between $d_{xz}$ ($d_{yz}$) and $d_{z^2}$ 
orbitals. Thus, for a deeper understanding of the negative sign in Eq.~(2), one 
needs to find the mechanism that determines the phases of different orbital
parts in the wave functions before SOI is included in the Hamiltonian. 
It is beyond the scope of our present study.
\bibitem{norm}
In Fig.~2, the norm, $|\psi_{n\vec{k},\mu}|$, of an orbital character,
$\mu$, of a state, $\psi_{n\vec{k}}$, is defined as
\begin{equation*}
  \,\,\,\,\,\,
  |\psi_{n\vec{k},\mu}| \! = \!\! \left[\mathrm{Re} \!\! 
  \left[\sum_{\alpha} \!\! \int \!\! d^3\!r \, 
  e^{-i\vec{k}\cdot\vec{\tau}_\alpha} c^*_{n\vec{k},\alpha\mu} 
  \phi^*_{\alpha\mu}(\vec{r}\!-\!\vec{\tau}_\alpha)
  \psi_{n\vec{k}}(\vec{r}) \right]\right]^{\!\frac{1}{2}}
  \end{equation*}
for $\mu = s$, $p_x$, $p_z$, $d_{z^2}$, $d_{xz}$, etc. Here $n$ is the 
band index, $\vec{k}$ is the wavevector, Re[$\cdots$] denotes the real 
part of a complex number, asterisks denote complex conjugates,
$\vec{\tau}_\alpha$ is the position of the $\alpha$th atom in the supercell, 
$\phi_{\alpha\mu}(\vec{r})$ is the pseudoatomic orbital of the orbital 
character, $\mu$, of the $\alpha$th atom, and $c_{n\vec{k},\alpha\mu}$ is 
the expansion coefficient of $\psi_{n\vec{k}}(\vec{r})$ using the 
pseudoatomic orbitals, i.e.,
\begin{equation*}
   \psi_{n\vec{k}}(\vec{r}) = \sum_{\alpha\mu\vec{R}}
   c_{n\vec{k},\alpha\mu} e^{i\vec{k}\cdot(\vec{\tau}_\alpha+\vec{R})}
   \phi_{\alpha\mu}(\vec{r}-\vec{\tau}_\alpha-\vec{R}),
\end{equation*}
where $\vec{R}$ is the in-plane lattice vector of the Au or Ag (111) 
surface. With this definition, the norm of the orbital character of a 
state satisfies $ \sum_{\mu} |\psi_{n\vec{k},\mu}|^2 = 1$, where the sum 
is over all orbital characters. In Figs.~2(a) and (b), the norm, 
$|\psi_{n\vec{k},\mathrm{all}\, d}|$, of all $d$-orbital characters, is 
defined as 
\begin{equation*}
|\psi_{n\vec{k},\mathrm{all}\, d}| = \sqrt{ \sum_\mu |\psi_{n\vec{k},\mu}|^2},
\end{equation*}
where the sum is over five $d$-orbital characters.
\end{thebibliography}

\end{document}